\documentclass{article}

\pdfoutput=1 

\usepackage{natbib}

\usepackage[utf8]{inputenc}

\usepackage{url}
\usepackage{mathtools}
\usepackage{authblk}

\usepackage[margin=1in]{geometry}

\newcommand{\rcode}[1]{\normalfont\texttt{#1}}

\title{Prediction approaches for partly missing multi-omics covariate data: A literature review and an empirical comparison study}
\author[1,*]{Roman Hornung}
\author[1]{Frederik Ludwigs}
\author[1,2,3,4]{Jonas Hagenberg}
\author[1]{Anne-Laure Boulesteix}
\affil[1]{Institute for Medical Information Processing, Biometry and Epidemiology, University of Munich, Munich, Germany}
\affil[2]{Department of Translational Research in Psychiatry, Max Planck Institute of Psychiatry, Munich, Germany}
\affil[3]{Institute of Computational Biology, Helmholtz Zentrum Mu\"unchen, Neuherberg, Germany}
\affil[4]{International Max Planck Research School for Translational Psychiatry, Munich, Germany}
\affil[*]{Corresponding author: Roman Hornung, hornung@ibe.med.uni-muenchen.de}

\begin{document}

\maketitle

\begin{abstract} 
As the availability of omics data has increased in the last few years, more multi-omics data have been generated, that is, high-dimensional molecular data consisting of several types such as genomic, transcriptomic, or proteomic data, all obtained from the same patients. Such data lend themselves to being used as covariates in automatic outcome prediction because each omics type may contribute unique information, possibly improving predictions compared to using only one omics data type. Frequently, however, in the training data and the data to which automatic prediction rules should be applied, the test data, the different omics data types are not available for all patients. We refer to this type of data as block-wise missing multi-omics data. First, we provide a literature review on existing prediction methods applicable to such data. Subsequently, using a collection of 13 publicly available multi-omics data sets, we compare the predictive performances of several of these approaches for different block-wise missingness patterns. Finally, we discuss the results of this empirical comparison study and draw some tentative conclusions.
\end{abstract}

\section{Introduction}
\label{sec:introduction}

The generation of various types of omics data is becoming increasingly rapid and cost-effective. As a consequence, there are more so-called multi-omics data becoming available, that is, high-dimensional molecular data of several types such as genomic, transcriptomic, or proteomic data measured for the same patients. In the last few years, several approaches to use these data for patient outcome prediction have been developed (see \citet{Hornung:2019} for an extensive literature review). Nevertheless, doubts have recently emerged as to whether there is benefit to using multi-omics data over simple clinical models \citep{Herrmann:2020}.

Regardless of their usefulness for prediction, multi-omics data from different sources that are used for the same prediction problem, for various reasons, often do not feature the exact same types of data. Most importantly, the data for which predictions should be obtained, that is, the test data, often do not contain the same data types as the data available for obtaining the prediction rule, that is, the training data \citep{Krautenbacher:2019}. The training data is also frequently composed of subsets originating from different sources (e.g. different hospitals) that frequently consist of different combinations of omics data types, as mentioned previously. When focusing on the collection of all omics data types available in at least one of the observations and considering data types not available for the different observations as missing, we can concatenate the data associated with all observations to obtain a large data set with partly missing data (cf.\ Figure \ref{fig:patterns} for illustrative examples). In the following, data concatenated in this form will be referred to as block-wise missing multi-omics data, where the different omics data types will be denoted as \lq\lq blocks''. The groups of observations in the data set that share the same combinations of observed data types will be denoted as \lq\lq subsets'' for simplicity. Note that in addition to the omics data, there are often clinical covariates (e.g. age or disease stage) available in practice, which usually contain plenty of predictive information. In this paper, we assume that the clinical covariates are always available for all patients. This data will be referred to as the \lq\lq clinical block'' in the following.

We begin with a presentation of the current state-of-the-art prediction approaches applicable to block-wise missing multi-omics data, followed by an empirical benchmark comparison study of the performance of some approaches. For this study we used a large collection of publicly available multi-omics data sets, for which missing data was generated artificially. As a response variable, we used the presence versus absence of a TP53 mutation. While it is not clinically meaningful to derive a model that predicts the presence of TP53 mutations, these mutations have been associated with poor clinical outcomes \citep{Wang:2017}. Against this background, we use TP53 as a surrogate for a phenotypic outcome in our benchmark study.

\section{Existing prediction approaches for partly missing multi-omics covariate data}
\label{sec:existing}

Existing methods that can handle block-wise missing data can be broadly characterised into three categories: naive methods, imputation methods to be used before prediction modelling, and methods that deal with the missingness structure within the prediction modelling process. In the following, we will describe these methods, where Section \ref{sec:approaches} will provide greater detail on those methods that are included in our empirical comparison study.

\subsection{Naive methods}
\label{sec:naive}

The simplest method is the complete case approach. The training is performed on the subset of all observations with complete observations. To increase the number of training observations, it makes sense to only use the blocks that are also contained in the test data. This method has the major disadvantage that it leads to discarding a potentially important part of the training observations. In extreme cases, there may not even be any patients with complete data. For example, one study center may have collected transcriptomic and proteomic data only, while the other center has collected transcriptomic and genomic data only; none of the patients are a complete case with respect to the combination of transcriptomic, proteomic, and genomic data.

Another naive method is the single block approach, where a model is trained only on one block that is available in both the training and test data. Here it is efficient to use the block associated with the greatest cross-validated or out-of-bag predictive performance. This method has the major disadvantage that it leads to discarding a potentially important part of the predictor variables. The complete case approach and the naive method share the advantage that, for training, any prediction method suitable for high-dimensional covariate data can be used.

\subsection{Imputation-based methods}
\label{sec:impute}

In the simplest form, imputation methods can be used to impute the missing values in the same ways as performed in conventional missing data imputations. Here, missForest \citep{Stekhoven:2011} or a semi-supervised learning approach \citep{Lan:2022} may be used. TOBMI \citep{Dong:2019} also uses a similar approach but makes use of the block structure to define the similarity between observations. With the latter methods, as with those presented in the last subsection, any prediction method suitable for high-dimensional covariate can be used.

Several methods use matrix completion algorithms to impute missing values. In \citet{Thung:2014}, first, the training data is reduced in size by removing noisy and redundant covariates and selecting observations that well represent the observations in the test data. Second, after removing the same covariates from the test data, the missing data in the combined training and test data are imputed. \citet{Cai:2016} assume an approximately low rank matrix for block-wise missing data and propose a structured matrix completion algorithm. This is the basis for \citet{Linder:2019} who also allow for missing values in individual covariates in addition to the block-wise missingness structure.

Other approaches for imputation include \citet{Zhu:2020} who assume that the data are realisations from exponential families and estimate the parameters of the missing values by taking relations within and between omic types into account. In the multiple block-wise imputation approach \citep{Xue:2021}, the data is divided into disjoint groups based on the missingness pattern and missing values are imputed multiple times. The imputed data sets are used to generate estimation equations and the different estimators are combined to yield one prediction. MI-GAN \citep{Dai:2021} learns generative adversarial networks for the different missingness patterns to multiply impute the missing values. Multi-Block Data-Driven sparse partial least squares (mdd-sPLS) \citep{Lorenzo:2019} first performs partial least squares (PLS) regression per block. The predictors from the individual PLS models are then combined, missing values are mean imputed, and then the PLS procedure is applied in an expectation maximisation fashion to generate better imputations.

Some methods make use of latent structure for the imputation. \citet{Zhang:2020} estimate a factor model that is then used to impute missing values. \citet{Yang:2020} first learn the pairwise mappings between each data type $A$ and the remaining data types using autoencoders. Second, these pairwise mappings are regularised to be consistent with each other, each pairwise mapping is applied to $A$ imputing the missing values, and finally the imputations obtained using the different pairwise mappings are averaged. Note while this method was not designed for multi-omics data, it can still be used for such data.

\citet{Hieke:2016} build different penalised regression models and use the predictions from one model as the offset for the next model. For observations that do not feature predictions from one model due to missing values, only the offset is imputed and not the missing values themselves. The same idea is used and generalised in priority-LASSO-impute (pL-imp (available)) \citep{Hagenberg:2020}. Sequential LASSO models based on one block and the predictions from the previous model as offset are fitted. Missing offsets are imputed using all available data from the other blocks.

\subsection{Methods that deal with the missingness pattern}
\label{sec:pattern}

A different approach is to divide the data into groups based on the missingness pattern and then deal with this structure without imputing missing values. One of the earliest methods of this kind was devised by \citet{Ingalhalikar:2012} who divided the data into subsets so that every subset has complete observations for a specific combination of blocks. On each of these subsets a model is trained and the predictions are weighted inversely by their expected error. A similar approach are multi-source random forests, described and evaluated in a smaller-scale empirical study  \citep{Ludwigs:2020}. With this method, separate random forests are learned on subsets of the data where each contains observations that feature a particular combination of blocks. Similar to \citet{Ingalhalikar:2012}, the predictions of these forests are weighted by their out-of-bag prediction performance \citep{Breiman:2001}. Originally, multi-source forests were designed for a different purpose than prediction using partly missing multi-omics covariate data, see Section \ref{sec:approaches} for details. A similar idea is used in \citet{Krautenbacher:2019} where a model is trained on each block.

The methods described in the previous paragraph learn models independently on subsets of the training data. There are, however, also several methods which share information in the learning of models obtained on different subsets of the data. This has the advantage of increasing the robustness of the individual models. \citet{Yuan:2012} proposed the incomplete Multi-Source Feature (iMSF) learning method. First, the data are divided into disjoint subsets, where each subset contains observations that feature a particular combination of blocks (and no further blocks). Subsequently, a regularised regression model is fitted on every subset. These models are constrained in such a way that in all models that share a block, the same covariates within this block have to be selected. However, the coefficients for one block do not need to have the same values. In contrast to the methods discussed in the previous paragraph, for iMSF, it is not necessary to apply different models to the test data and obtain a weighted prediction. Instead, to obtain predictions the model with the correct block combination is applied. For example, suppose the training data featured blocks $A$, $B$, and $C$. In this case, using the training data, a separate model is learned using each possible combination, that is, using the following combinations: $\{A, B\}$, $\{A, C\}$, $\{B, C\}$, and $\{A, B, C\}$. Suppose now that a test observation is missing the block $B$, thus it only features the blocks $A$ and $C$. In that case, to obtain a prediction, the model learned on the combination $\{A, C\}$ is applied to the test observation. 

The iMFS method was developed further by \citet{Xiang:2014} to the incomplete Source-Feature Selection (iSFS) model. Again, different subsets are formed based on the missingness patterns, but in contrast to for iMFS these subsets are overlapping. Each subset is defined as all observations that feature all blocks from a particular combination of blocks, but potentially also other blocks. For example, if there are three blocks $A$, $B$, and $C$, the subset corresponding to the combination $A$ with $C$ contains all observations that feature only $A$ and $C$, but also all observations that feature $B$ in addition to $A$ and $C$. This procedure has the advantage that more observations are considered per combination compared to in the case of iMFS. Moreover, in contrast to iMFS, not a separate coefficient vector for each combination is learned. Instead, a common coefficient vector is learned for the variables from all blocks and to obtain the coefficient vectors for the individual combinations, the entries in the common coefficient vector are multiplied by block- and combination-specific weights. This has the advantage that less parameters need to be estimated for the iSFS, which should lead to more stable models. \citet{Lan:2021} reformulate the iMSF method as a multi-task learning problem.

\citet{Liu:2017} again divide the observations into subsets based on the missingness pattern in a similar way to iMSF, where, however, the available data are exploited better than for iMSF. Then a hypergraph is learned on every subset that also incorporates high-order relationships. The different hypergraphs are combined and used to train a classifier. In \citet{Dong:2021}, a similar approach is used. In contrast to \citet{Liu:2017}, the hypergraphs are learned on a low-rank representation of the data. Afterwards, for every block, a support vector machine classifier is trained and the predictions are combined. Another method working with hypergraphs is Heterogeneous Graph-based Multimodal Fusion \citep{Chen:2020}. The data are first divided into subsets based on the missingness pattern and heterogeneous hypernode graphs are constructed. On every graph, the relationships between different data types are learned and this information is used to construct a new hypergraph. Then, the interactions between the different missingness patterns are learned and the information of the different data types are fused into one embedding.

The methods by \citet{Chen:2020}, \citet{Ingalhalikar:2012}, and \citet{Lan:2021} were not designed for the multi-omics case. However, they can still be applied to such data.

\section{Empirical comparison study of approaches for partly missing multi-omics covariate data}
\label{sec:empcompstudy}

\subsection{Neutrality disclosure}
Our study is intended as \lq\lq neutral'' in the sense that we are focusing on the comparison rather than promoting a particular new method \citep{boulesteix2017necessity,boulesteix2017towards}. However, we are not equally familiar with all methods. In particular, two of the included methods, multi-source random forests and priority-LASSO-impute, were developed by some of the authors of this paper. While our familiarity with these methods helped us to set them up appropriately, we were committed to providing a fair comparison, that is, we neither spent more efforts to optimise these two methods than the other methods nor did we design the study to favour one or the other method.

\subsection{Design of the comparison study}
\label{sec:design}

\subsubsection{Compared approaches and their configurations}
\label{sec:approaches}

In this subsection, we provide greater detail on the compared approaches and their specific configurations in the empirical study. When selecting these approaches out of the methods described in Section \ref{sec:existing} we ensured that at least two methods from each of the three categories described in Sections \ref{sec:naive} to \ref{sec:pattern} were included. Moreover, we selected only methods that are either implemented in publicly available R packages or could be implemented with reasonable effort.

The following approaches were considered in the study: complete case approach, single block approach, imputation based on TOBMI, mdd-sPLS, block-wise random forest, multi-source random forest, and priority-LASSO-impute. As described in Section \ref{sec:naive}, the first two of these are naive approaches. These served as a baseline against which the other more sophisticated  methods were compared. Three of the other methods were imputation-based (imputation based on TOBMI and mdd-sPLS, as well as priority-LASSO-impute, see Section \ref{sec:impute}) and two methods deal with the missingness pattern without imputation procedure (block-wise random forest and multi-source random forest, see Section ref{sec:pattern}). In the following, instead of ``block-wise missing multi-omics training data'' and ``block-wise missing multi-omics test data'' the terms ``training data'' and ``test data'' will be used for simplicity.

\paragraph{Complete case approach (\rcode{ComplcRF})}
In conventional complete case analysis, all observations that contain missing values are removed. However, in the presence of block-wise missingness this would often be impossible or ineffective because there may be none or very few observations with all blocks observed. Therefore, to increase the number of training observations with no missing data, we first removed all those blocks from the training data that did not occur in the test data. Subsequently, we removed all observations with missing values from the training data. Lastly, a random forest prediction rule \citep{Breiman:2001} was constructed on the training data. Note that with the complete case approach, as in the majority of the approaches described below, the prediction rule can only be trained after knowing the missing data structure of the test data. No tuning parameter optimisation was performed for the random forests: the tuning parameter values were set to the default values of the R package 'randomForestSRC' (version 2.9.2). For example, the numbers of covariates sampled for each split $mtry$ were set to the square roots of the numbers of covariates and 500 trees were constructed per forest. We proceed in the same way in the cases of the random forests constructed for the single block approach, the imputation approach based on TOBMI, and the block-wise random forests (all described below).

\paragraph{Single block approach (\rcode{SingleBlRF})}
Even though the available data set may include several blocks, training may be conducted using only individual blocks. This may be advantageous in situations in which a single block carries most of the available predictive information or in which the predictive information contained in the different blocks is redundant. In the first step of the single block approach, we removed all blocks not featured in the test data from the training data. Subsequently, we trained a random forest on each block and, using the out-of-bag predictions \citep{Breiman:2001} and the true values of the response variable, calculated performance measure values of these random forests. The area under the receiver operating characteristic (AUC) was used as a performance measure. Finally, a random forest was trained on the block associated with the largest calculated AUC value.

\paragraph{Imputation approach based on TOBMI (\rcode{ImpRF})}
A more sophisticated approach to dealing with missing data than the complete case approach is imputing the missing values using a data-driven procedure. An important advantage of the imputation approach over the complete cases approach is that no observations have to be excluded when training the prediction rule.

TOBMI was designed for situations with two (omics) blocks A and B, where block A is available for all observations and block B is missing for part of the observations. First, the Mahalanobis distance matrix $M$ between all observations is calculated, where importantly only the measurements from block A are used. Subsequently, the data of each observation $i$ with missing data from block B are imputed in the following way: 1) Determine the $k$ observations in the subset of the observations with measurements in block B that are closest to $i$ according to $M$ (note again that the latter distance matrix was obtained only using block A). These $k$ observations likely behave similar to $i$ and thus are used for imputing the missing values of $i$ in the second step; 2) Impute each missing block B value in $i$ by its weighted mean across the $k$ nearest neighbours determined in step 1). As weights, the reciprocals of the Mahalanobis distances from $i$ are used. For $k$, the (rounded down) square root of the number of observations with measurements in block A and B is used.

Since, as described in the previous paragraph, TOBMI is not applicable to general block-wise missingness patterns, we proceeded as follows to impute the training data in our comparison study. We first concatenated all those blocks that featured no missing values for any of the observations. Subsequently, we used TOBMI repeatedly, each time for imputing the values of a different block with missing values. Here, the concatenation of the blocks without missing values took the role of 'block A' from the previous paragraph and the block to impute at the current repetition took the role of 'block B'. The imputation was performed across subsets. Note that with this procedure, there needs to be at least one block that is available for all observations because we need at least one block that can be used to calculate the Mahalanobis distances between all observations (i.e. Matrix $M$). This should be fulfilled in practice in most cases because usually clinical information is available for all patients.

Subsequently, we removed all blocks in the training data not available in the test data set and constructed a random forest using the training data processed in this way.

\paragraph{Multi-block Data-Driven sparse PLS (mdd-sPLS) (\rcode{MddsPLS})}
Like the imputation approach described above, mdd-sPLS, introduced by \citet{Lorenzo:2019}, involves imputing the data set. Unlike the imputation approach based on TOBMI, not only the training data set is imputed, but also the test data set, where these imputations are performed separately. After imputation there are no missing values in training and test data, which is why all blocks can be used for prediction. The mdd-sPLS algorithm is a complex method based on the partial least squares (PLS) technique. Put simply, mdd-sPLS performs PLS type procedures separately on the blocks, combines the information afterward, and predicts the outcome. The missing values in training and test data of those covariates that are used by mdd-sPLS in prediction are imputed using additional PLS-based models. The missing values are imputed using the outcome and the other covariates, respectively. These two steps, building the prediction model and imputing missing values, are repeated until convergence of the involved latent variables. 

We used 10-fold cross-validation to determine the optimal regularisation parameter for the correlation matrices, performing a grid search on 10 values. Moreover, we used one component for the involved matrix decomposition and inversely weighted the components per block by the number of selected covariates per block. See \citet{Lorenzo:2019} for details.

\paragraph{Block-wise random forest (\rcode{BlwRF})}
Unlike the approaches described above, the method introduced by \citet{Krautenbacher:2018}, which we denote ``block-wise random forest'', uses all available measurements from the training data set without performing imputation. A separate random forest is constructed for each block; in each case, all observations that feature measurements for the corresponding block in the training data are used. To obtain predictions for the test data, first the corresponding random forests are applied to each block available in the test data, and in each case a predicted probability for $Y = 2$ is calculated, where $Y \in \{1, 2\}$ describes the binary response. Second, a weighted average of the predicted probabilities obtained for each test data block is calculated with weights proportional to the out-of-bag calculated AUC values of the corresponding random forests. Note that, in general, other weights can be used as well, for example weights based on performance measures other than the AUC or equal weights for all blocks, where the latter would correspond to an unweighted average.

\paragraph{Multi-source random forest (\rcode{MultisRF})}
This method was originally designed for situations with several training data sets with varying, potentially overlapping covariate sets, but a common response variable. The difference to the situations considered in this paper is that with the situations originally targeted by \rcode{MultisRF}, there are no blocks of missing data, but the missingness patterns are more diffuse with different covariates missing for each data set (or 'subset' in the terminology used in this paper). The test data include a subset of the covariate sets available in the training data sets.

For training the prediction rule a standard random forest is constructed on each training data set. For prediction, first, each tree in these forests is pruned in the following way: Starting with the first split that divides the full (bootstrapped or subsampled) data set, each branch in the tree is followed and for each encountered split it is checked whether that split uses a covariate available in the test data and if not, the branch is cut. This ensures that the trees in the forests only use covariates that are available in the test data. Subsequently, similarly to in the case of \citet{Krautenbacher:2018}, each forest is applied to the test data and each prediction is weighted proportionally to the out-of-bag AUC value of the respective forest.

The procedure described above can be applied as is in prediction using partly missing multi-omics covariate data. Here, for the data sets with varying covariate sets subsets of the data are used, where each subset contains observations that feature a particular combination of blocks. This approach is implemented in the R package 'multisForest' available on GitHub (\url{https://github.com/RomanHornung/multisForest}, version 0.1.0). Due to the comparably high computational burden associated with this approach we used only 250 trees per forest instead of 500 trees as in the cases of the other random forest-based methods. For the remaining tuning parameter values we used the default values from the R package 'randomForestSRC' (version 2.9.2).

\paragraph{priority-LASSO-impute (pL-imp (available)) (\rcode{PrLasso})}
The pL-imp (available) algorithm, developed by \citet{Hagenberg:2020}, is an extension of the multi-omics prediction method priority-LASSO \citep{Klau:2018} for block-wise missing multi-omics data. The method is available in the R package 'prioritylasso' on GitHub (\url{https://github.com/jonas-hag/prioritylasso}) \citep{Klau:2020}. Priority-LASSO is a method based on the Least Absolute Shrinkage and Selection Operator (LASSO) \citep{Tibshirani:1996} that allows researchers to specify a priority ranking of the blocks. A priority ranking is often given because some blocks are more established or easier to obtain than others. The first step of obtaining the priority-LASSO prediction rule is to fit a LASSO model using only the covariates in the block with highest priority. A LASSO model is again fit in the second step, however, this time using only the covariates in the block with second-highest priority using the linear predictor from the LASSO model fitted in the first step as an offset in the model equation. By including the latter offset, only that part of the predictive information contained in the block with second-highest priority that is not contained in the block with highest priority is used. This process is continued iteratively for all blocks in the order of their priority, this way obtaining estimated model coefficients for all covariates.

In the case of block-wise missing multi-omics data, this estimation scheme would not be applicable because in each step the offsets would be available only for those observations for which the respective block is available. \citet{Hagenberg:2020} considered several solutions to this issue. In this paper, only one of these will be used, pL-imp (available), because it showed good results in \citet{Hagenberg:2020} and also seems to be the most promising from a conceptual point of view. With this approach, the missing offsets are imputed using a LASSO model. To simplify, in each step for every subset this approach first learns a LASSO model using observations for which the offsets are available. In this LASSO model, the offsets are the response variables and the other blocks contain the covariates. The fitted model is then used to predict the missing offsets.

The pL-imp (available) algorithm was not designed to handle missing blocks in the test data. In order to deal with this issue, we excluded all blocks that were not available in the test data before fitting the model to the training data. Moreover, the pL-imp (available) algorithm, similar to the original priority-LASSO algorithm, requires the user to provide a priority ranking of the available blocks. As no useful biological information was available for the data sets considered in the comparison study performed for this paper, the priority rankings were determined in the following way: 1) Fitting a LASSO model to each block and estimating the deviance associated with each model using 5-fold cross-validation (CV); 2) Assigning the highest priority to the block associated with the lowest cross-validated deviance value, the second highest priority to the block associated with the second-lowest cross-validated deviance value, and so on; in cases in which two or more blocks were associated with the same cross-validated deviance value, the priority order between these blocks was assigned randomly.

The shrinkage parameters in the LASSO models involved in the priority-LASSO and priority-LASSO-impute estimation procedures were determined using grid search and 10-fold CV.

\subsubsection{Data}
\label{sec:data}
The data material consists of 13 publicly available multi-omics data sets from The Cancer Genome Atlas (TCGA) project \citep{Weinstein:2013}. These data are a subset of 21 data sets previously used in \citet{Hornung:2019}. From these 21 data sets, 18 contained all four omics blocks that were considered as covariates (see below), the clinical block and the response variable, that is, the presence versus absence of the TP53 mutation. From the remaining 18 data sets, we removed imbalanced data sets for which the smaller response variable class was represented by less than 15\% of the observations. This resulted in 13 data sets for use in the comparison study. The covariates consisted of the following four blocks: clinical block, copy number variation block, miRNA block, and RNA block. Table \ref{tab:data} gives an overview on the used data sets.

\begin{table}[ht]
\centering
\caption{Overview on the data sets. The second column shows the number of observations. The third column shows the proportion of observations with TP53 mutation. The fourth to the seventh column show the numbers of covariates in the respective blocks (clin: clinical covariates, cnv: copy-number variation, mirna: miRNA, rna: RNA).} 
\begin{tabular}{lrrrrrr}
  \hline
label & n & prop.\ TP53 & clin & cnv & mirna & rna \\ 
  \hline
BLCA & 310 & 0.49 &   4 & 57964 & 825 & 23081 \\ 
  BRCA & 863 & 0.34 &   8 & 57964 & 835 & 22694 \\ 
  COAD & 350 & 0.56 &   5 & 57964 & 802 & 22210 \\ 
  ESCA & 121 & 0.83 &   4 & 57964 & 763 & 25494 \\ 
  HNSC & 411 & 0.69 &   5 & 57964 & 793 & 21520 \\ 
  LGG & 454 & 0.46 &   3 & 57964 & 645 & 22297 \\ 
  LIHC & 298 & 0.29 &   4 & 57964 & 776 & 20994 \\ 
  LUAD & 424 & 0.49 &   6 & 57964 & 799 & 23681 \\ 
  LUSC & 365 & 0.85 &   7 & 57964 & 895 & 23524 \\ 
  PAAD & 142 & 0.63 &   4 & 57964 & 612 & 22348 \\ 
  SARC & 183 & 0.36 &   2 & 57964 & 778 & 22842 \\ 
  STAD & 284 & 0.47 &   6 & 57967 & 787 & 26027 \\ 
  UCEC & 503 & 0.36 &   3 & 57447 & 866 & 23978 \\ 
   \hline
\end{tabular}
\label{tab:data}
\end{table}

\subsubsection{Generation of block-wise missingness and performance evaluation}
\label{sec:patterns}
\begin{figure}
  \centering
\includegraphics[width = 0.7 \textwidth]{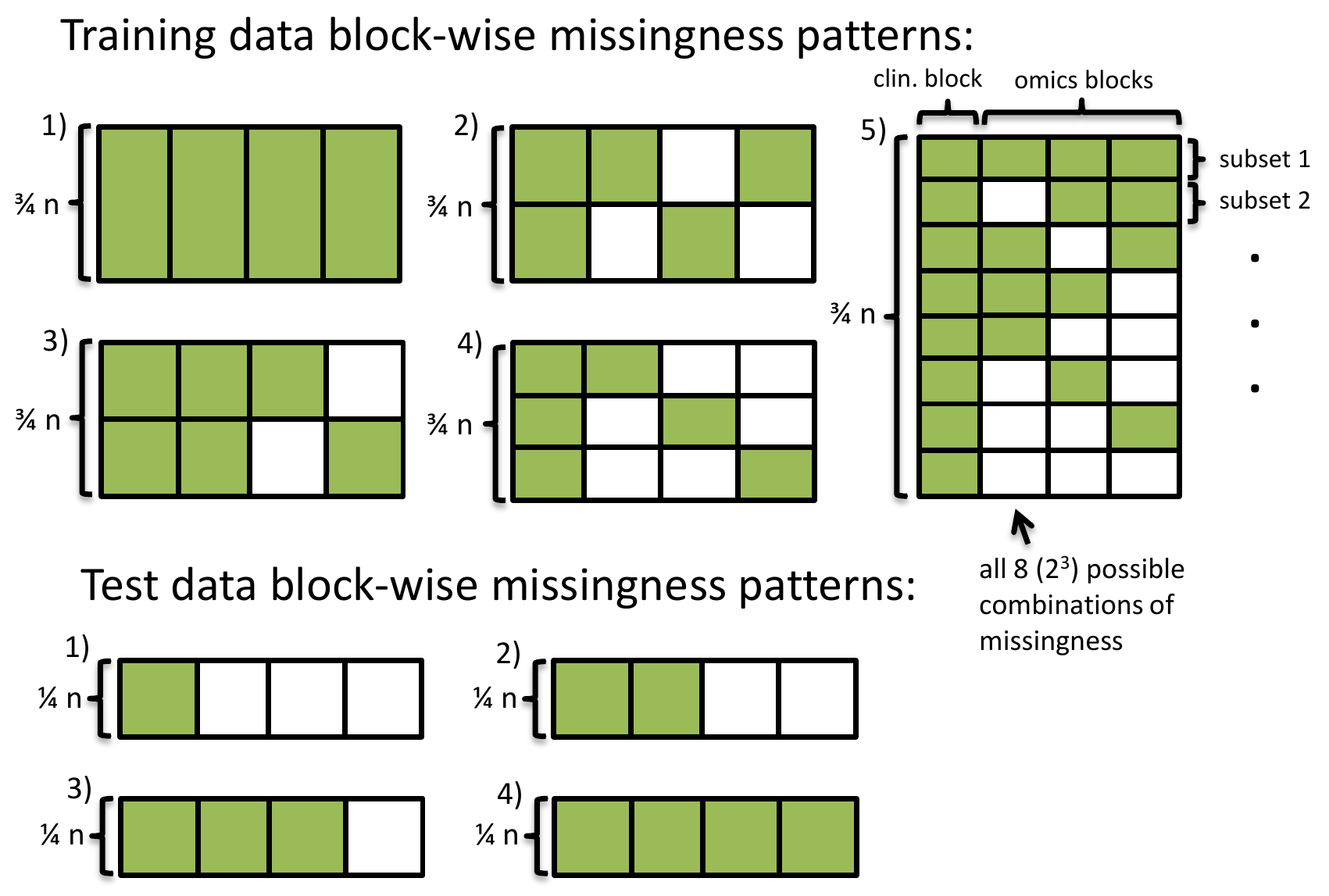}
\caption{Block-wise missingness patterns in training and test data. Coloured/empty boxes indicate that the respective blocks are present/absent. The first column always represents the clinical block.}
 \label{fig:patterns}
\end{figure}

Block-wise missingness patterns are generated by randomly deleting parts of the data sets. First, the data sets are split into training and test data in the ratio 3:1. Second, as described in more detail in the next paragraph, the block-wise missingness patterns are induced separately in training and test data, where there are five different patterns for the training data sets and four for the test data sets, see Figure \ref{fig:patterns}. In the following, training data block-wise missingness patterns will be abbreviated as \lq\lq trbmp'' and test data block-wise missingness patterns as \lq\lq tebmp'', respectively. As is evident from Figure \ref{fig:patterns}, each of the trbmps consists of either one, two, three, or eight subsets of observations. For each combination of data set, trbmp, and tebmp, the above procedure is repeated five times.

Following each division into training and test data, the subset memberships of the training observations are assigned randomly and the subsets are of equal size for each training data set (note again that the data are split into training and test data in the ratio 3:1). We induce the missingness patterns after random permutation of the omics blocks, where a different permutation is used for each repetition. This is performed separately for the training data and the test data. First, the omics blocks are permuted randomly (the clinical block always stays at the first position). Second, values in the data matrix are deleted according to the respective considered block-wise missingness pattern (Figure \ref{fig:patterns}). After having performed these steps separately for the training and test data, the blocks are re-ordered again to have the original ordering to ensure that the block ordering is the same in training and test data. Without the random permutation of the omics blocks described above, each block would have been observed with unequal frequency for the different missingness patterns. This would have made it impossible to tell whether differences in the results observed for different missingness patterns (trbmps and tebmps) are actually due to the missingness patterns or the fact that specific influential blocks are featured to different degrees in the missingness patterns. The permutation procedure ensures that different blocks are missing in the subsets for different repetitions even when considering the same trmps and tebmps. For example, consider trbmp 2 and tebmp 3; for the first division into training and test data, the first set in the training data might include RNA and miRNA data and the second set only mutation data, whereas the test data might include RNA and mutation, but no miRNA data. For the second division, the first set in the training data might include only mutation and miRNA data and the second set only RNA data, while the test data may include mutation and miRNA data, but no RNA data.

For each division, the methods described in Section \ref{sec:approaches} are learned on the training data, subsequently applied to the test data, and the performance on the test data measured according to three metrics: the Brier Score, the area under the receiver operating characteristic curve (AUC), and the accuracy. The Brier Score measures discrimination and calibration and is a proper scoring rule in the sense that it measures the accuracy of class probability predictions (e.g. probability for class B if there are two classes A and B). The accuracy is not a proper scoring rule because it does only evaluate the precision of class predictions, not class probability predictions. Thus, it uses less information from the predictions than the Brier Score. The AUC only measures discrimination. More precisely, it measures the models' abilities to order different subjects correctly in terms of their predicted probabilities. Note that a model can feature a high AUC value even though it is not calibrated well. This would, for example, be the case in a situation in which the predicted probabilities output by a model are systematically too small, but still observations with larger predicted probabilities for class B tend to feature class B much more often those with smaller predicted probabilities.

\subsubsection{Code availability}
\label{sec:code}
All R code written to perform and evaluate the analyses, including data pre-processing, is available on GitHub (\url{https://github.com/RomanHornung/bwm_article}). The pre-processed data sets are available as CSV files on the online open access repository figshare (\url{http://doi.org/10.6084/m9.figshare.20060201.v1}). For details on the pre-processing see \citet{Hornung:2019}.

\subsection{Results}
\label{sec:results}

As a preliminary remark, please note that some methods delivered few or no successful predictions for specific trbmps and tebmps. For example, \rcode{ComplcRF} is not applicable for tebmp 4 when considering trbmp 2, 3, or 4: For tebmp 4, the test data do not contain missing values, which is why none of the variables in the training data are removed (cf.\ Section \ref{sec:approaches}); because of this there are no complete cases in the training data for any of the trbmps 2, 3, or 4 (cf.\ Figure \ref{fig:patterns}) and the complete case approach \rcode{ComplcRF} is thus not applicable. \rcode{MultisRF} is not applicable for trbmp 1 because the R package 'multisForest' (version 0.1.0) implementing \rcode{MultisRF} does not allow training data sets without missing values. In addition, \rcode{PrLasso} lead to errors in rare cases. More precisely, ten of the 1300 repetitions performed in total for \rcode{PrLasso} resulted in an error, which all but one occurred for the data set ESCA. The remaining methods did deliver predictions in all cases. The frequencies of repetitions with missing results were 25.0 \%, 34.1 \%, and 0.8 \% for \rcode{ComplcRF}, \rcode{MultisRF}, and \rcode{PrLasso} respectively. We describe the reasons for unsuccessful predictions in full detail in Section A of the Supplementary Materials.

\subsubsection{Global performance comparison}
\label{sec:globperf}

\begin{figure}
\centering
\includegraphics[width=0.8\textwidth]{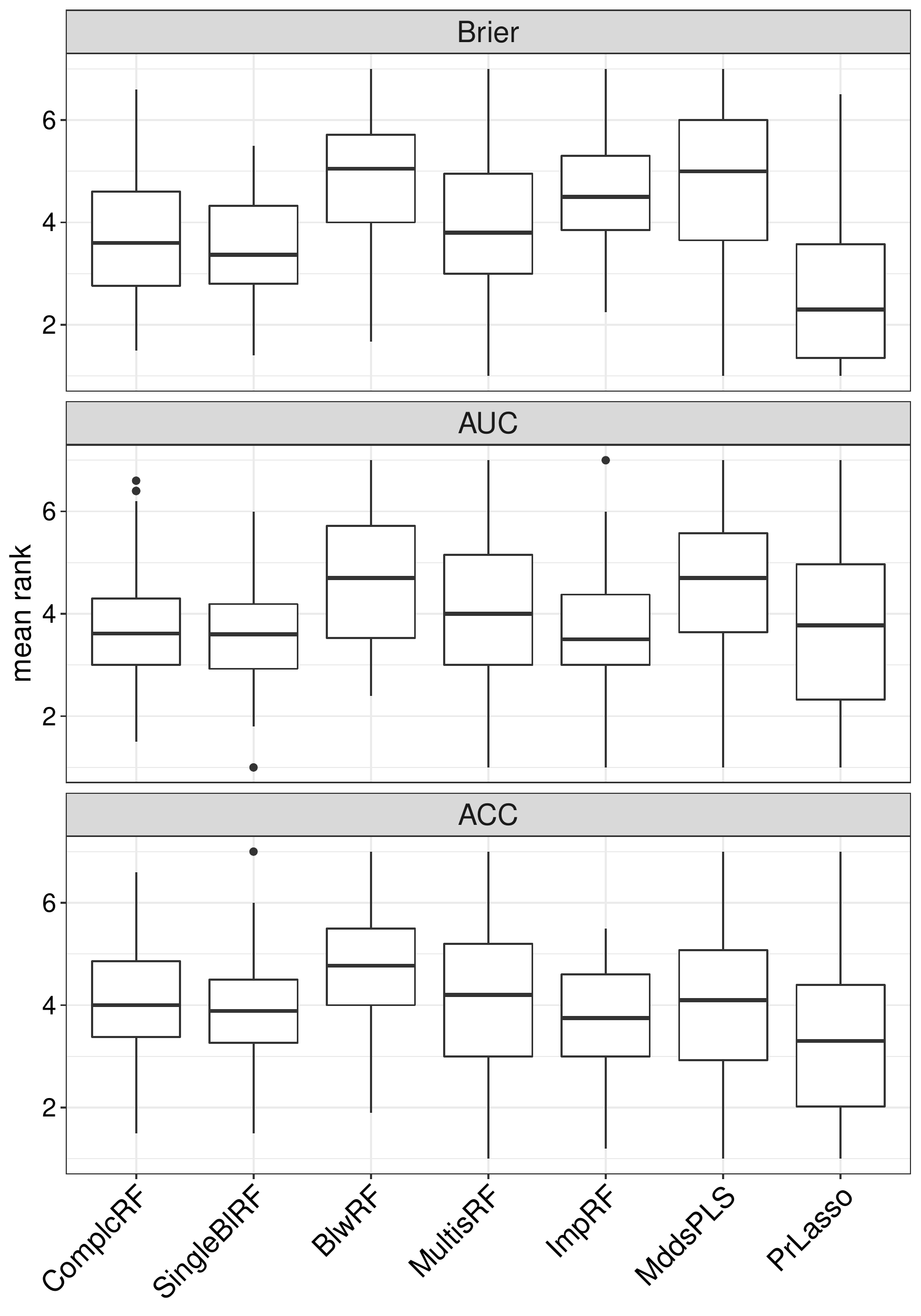}
\caption{Ranks each method achieved among the other methods in terms of the three considered performance metrics -- global performance. The ranks were computed for each combination of trbmp, tebmp, data set, and repetition, where only those repetitions were considered for which the results were available for all seven considered methods. The ranks were then averaged across the repetitions. Smaller ranks indicate a better performance.}
\label{fig:global_ranks}
\end{figure}

Figure \ref{fig:global_ranks} shows the ranks the methods achieved among each other, pooled across all trbmps, tebmps, and data sets. Note that here we only included those repetitions for which each of the seven considered methods delivered a result. If we would have included all available repetitions for each method, the comparison between the methods would not have been fair. This is because, as stated above, some methods did not deliver predictions for specific trmps and tebmps. Given that the predictive performance generally differed across different trbmps and tebmps, the comparison of the methods would have been confounded by the trmps and tebmps if all available results would have been used.

In general, the differences between the performances observed for the different methods are not very strong. For the Brier score and the accuracy \rcode{PrLasso} performed best, while for the AUC there is no clear winner among the methods. \rcode{BlwRF} was among the worst-performing methods for all three performance metrics, while for \rcode{MddsPLS} this was the case only for the Brier Score and the AUC. 

As stated above, in Figure \ref{fig:global_ranks}, we only included those repetitions for which results were available for all seven methods. Because of this, many repetitions were excluded. This was mainly due to \rcode{ComplcRF} and \rcode{MultisRF}. For example, as stated above, all repetitions with trbmp 1 were excluded because \rcode{MultisRF} did not deliver results for this method. Therefore, Supplementary Figures S1 and S2 show the results presented above, however, under the exclusion of \rcode{ComplcRF} and \rcode{MultisRF}, respectively. Excluding \rcode{ComplcRF} and \rcode{MultisRF} allowed us to consider more repetitions because we did not need to exclude all repetitions for which results were not available for \rcode{ComplcRF} and \rcode{MultisRF}, respectively. We do not see any notable differences in the results after the exclusion of \rcode{ComplcRF} and \rcode{MultisRF}, respectively.

Supplementary Figures S3, S4, and S5 show the raw values of the metrics, which confirm that the differences between the results obtained for the different methods are not strong.

\subsubsection{Performance separately by trbmp}
\label{sec:septrbmp}

\begin{figure}
\centering
\includegraphics[width=0.8\textwidth]{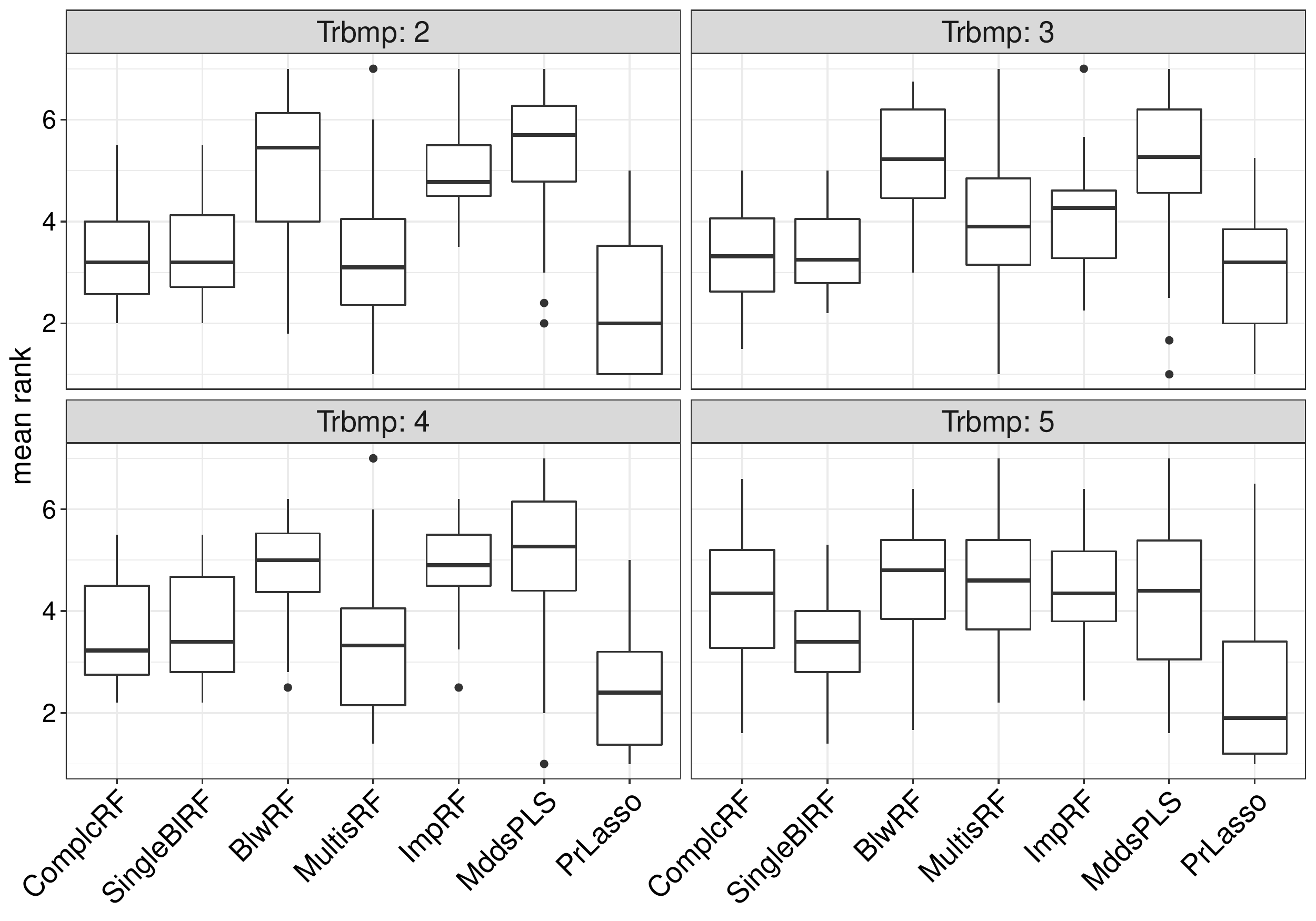}
\caption{Ranks each method achieved among the other methods in terms of the Brier score -- separately by trbmp. The ranks were computed for each combination of trbmp, tebmp, data set, and repetition, where only those repetitions were considered for which the results were available for all seven methods. The ranks were then averaged across the repetitions. Smaller ranks indicate a better performance.}
\label{fig:trbmp_Brier_ranks}
\end{figure}

Figure \ref{fig:trbmp_Brier_ranks} shows the ranks each method achieved with respect to the Brier score among the other methods separately by trbmp. For reasons of clarity, in the main paper, we do not present these results for all three performance metrics. The corresponding results obtained for the AUC and the accuracy are shown in Supplementary Figures S6 and S7. The reason why we decided on the Brier score was that it can be seen as the most important metric because it measures both discrimination and calibration, whereas the AUC and the accuracy each measure only one of these. In the following descriptions we will focus on the Brier score, but also describe differences observed for the other two performance metrics.

\rcode{PrLasso} performed best for all four trbmps shown in Figure \ref{fig:trbmp_Brier_ranks}. Note that trbmp 1 is not included here because \rcode{MultisRF} is not applicable for trbmp 1 and we again only considered repetitions for which results were available for all seven methods. For trbmps 2 to 5, \rcode{BlwRF} and \rcode{MddsPLS} were again among the worst-performing methods. For trbmps 2, 4, and 5, \rcode{ImpRF} was also among the worst-performing methods, but not for trbmp 3. The observation that \rcode{ImpRF} was not among the worst methods for trbmp 3 can likely be explained by a feature of \rcode{ImpRF}. As described in Section \ref{sec:approaches}, \rcode{ImpRF} uses the concatenation of those blocks that are observed for all observations to calculate the distance matrix that is used in the imputation. For trbmp 3, two blocks are available for all observations, whereas for the other trbmps this is the case for only one block (excluding trbmp 1 with no missing observations). Thus, for trbmp 3 the calculated distance matrix can be expected to better reflect the true distances between the observations, which would explain, why \rcode{ImpRF} performed better for trbmp 3 than for the other trbmps. However, we did not see this for the AUC and the accuracy (Supplementary Figures S6 and S7), where \rcode{ImpRF} generally performed better compared to the results obtained for the Brier score.

For trbmp 5, excluding \rcode{PrLasso} and \rcode{SingleBlRF}, all methods performed similarly poorly. \rcode{ComplcRF} likely performed worse for trbmp 5 than for the other trmps because the number of complete observations is much smaller for this trbmp. The likely reason \rcode{MultisRF} performed worse for trbmp 5 was that the subsets were mich smaller in size in comparison to the other trbmps. The predictive performance of the random forests learned on these small subsets probably suffered. Again, \rcode{ImpRF} performed better with respect to the AUC and the accuracy for trbmp 5.

The results obtained under the exclusion of \rcode{ComplcRF} and \rcode{MultisRF}, were again very similar (Supplementary Figures S8 to S13). The results obtained for trbmp 1 can only be studied by excluding \rcode{MultisRF} (Supplementary Figure S11) because this method was not applicable for trbmp 1. These results were similar to those obtained for trbmps 2, 3, and 4 expect that \rcode{BlwRF} clearly performed worst here. However, for the AUC \rcode{MdssPLS} performed similarly bad (Supplementary Figure S12). Note that, if there are no missing blocks in the training data (i.e. for trbmp 1), \rcode{ComplcRF} and \rcode{ImpRF} are identical.

In general, for the AUC and the accuracy, the differences in performance between the methods were smaller than for the Brier score (Supplementary Figures S6, S7, S9, S10, S12, and S13). The raw values of the performance metrics obtained for the analysis separated by trbmp are shown in Supplementary Figures S14 to S22.

\subsubsection{Performance separately by tebmp}
\label{sec:septebmp}

\begin{figure}
\centering
\includegraphics[width=0.8\textwidth]{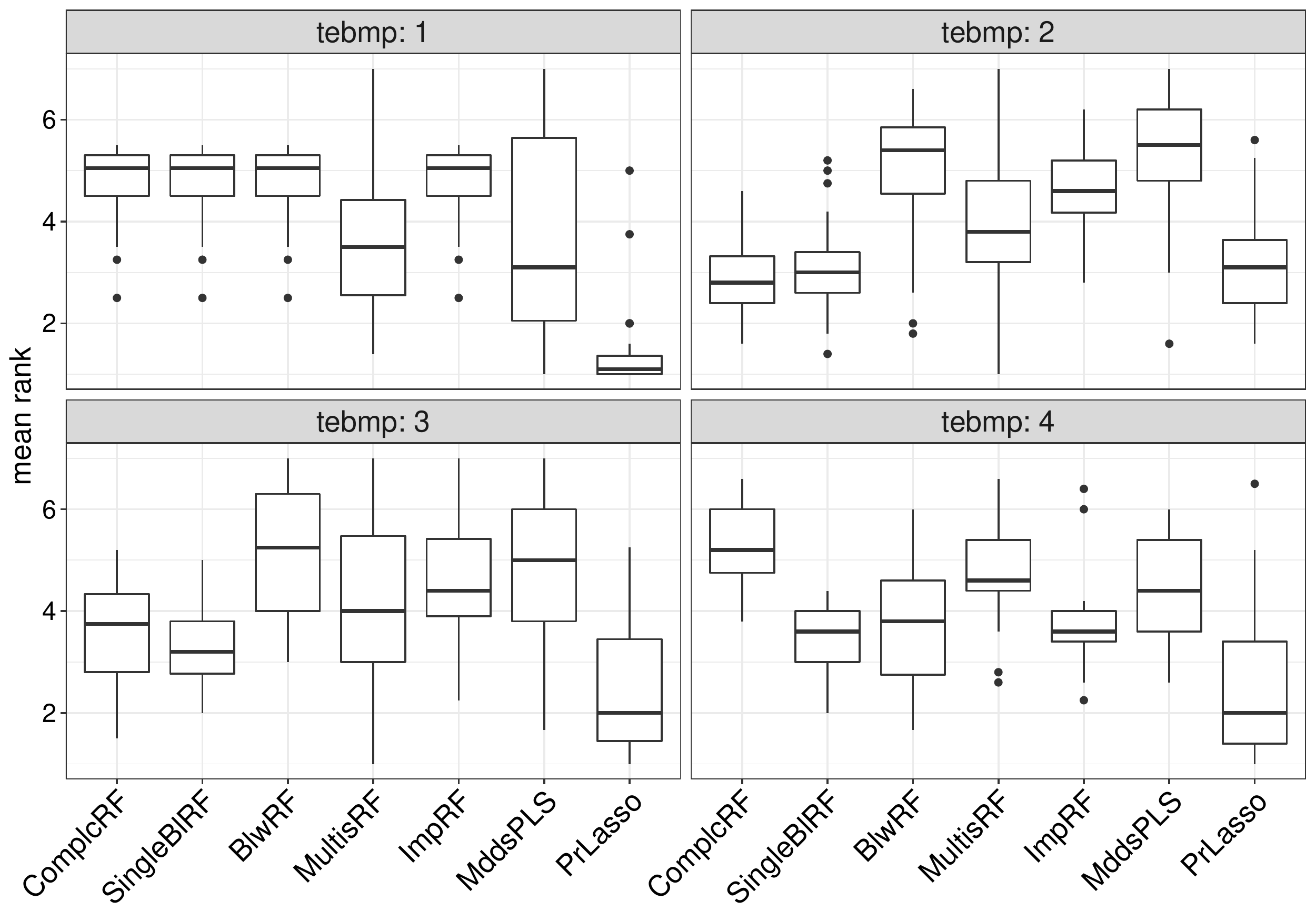}
\caption{Ranks each method achieved among the other methods in terms of the Brier score -- separately by tebmp. The ranks were computed for each combination of trbmp, tebmp, data set, and repetition, where only those repetitions were considered for which the results were available for all seven methods. The ranks were then averaged across the repetitions. Smaller ranks indicate a better performance.}
\label{fig:tebmp_Brier_ranks}
\end{figure}

For comparing the results observed for the different methods separately by tebmp we again focus on the Brier score (Figure \ref{fig:tebmp_Brier_ranks}). \rcode{PrLasso} again performed the best for all tebmps excluding tebmp 2, where \rcode{ComplcRF} and \rcode{SingleBlRF} performed similarly well. For \rcode{ComplcRF}, this good performance may be explainable by the fact that for tembmp 2 there are few blocks observed in the test data, which is why many blocks are removed from the training data which in turn increases the number of complete observations in the training data. Note, however, that for the AUC and the accuracy (Supplementary Figures S23 and S24), \rcode{PrLasso} was not clearly the best method. 

For tebmp 1 \rcode{ComplcRF}, \rcode{SingleBlRF}, \rcode{BlwRF}, and \rcode{ImpRF} all performed equally well. This can be explained by the fact that for this tebmp only the clinical block is available in the test data and for these methods all blocks not available in the test data are removed from the training data. Therefore, these four methods all function identically for tebmp 1 because they all construct standard random forests using only the clinical block. Note that for tebmp 1, \rcode{PrLasso} corresponds to a simple Lasso model fitted to the clinical block. Against this background it is interesting to see that \rcode{PrLasso} still performed better than these four random forest-based methods named above (not in the case of the AUC, see Supplementary Figure S23). This means that, when using the clinical block as covariate data, standard Lasso performs better than standard random forests. Tebmp 1 is also the only setting for which \rcode{MddsPLS} worked better than these four random forest-based methods named above. Given that tebmp 1 is the only tebmp for which all blocks except the clinical block are missing in the test data, the better performance of \rcode{MdssPLS} for tebmp 1 might be due to the fact that this method is the only one that imputes the missing blocks in the test data. For the AUC, however, \rcode{MdssPLS} performed worse than the four random forest-based methods named above also for tebmp 1; for this metric, \rcode{MultisRF} also performed worse.

When interpreting the results for the remaining tebmps, it must be considered that, in Figure \ref{fig:tebmp_Brier_ranks}, the results displayed for tebmp 4 are only those obtained for trbmp 5. In this figure, as before, we show only the results of those repetitions for which each of the seven considered methods delivered a result. As explained above, for tebmp 4, the results of either \rcode{ComplcRF} (for trbmps 2, 3, and 4) or \rcode{MultisRF} (for trbmp 1) were missing for all trbmps excluding trbmp 5. Therefore, when interpreting the results obtained for tebmp 4 we must resort to Supplementary Figure S25, which shows the corresponding results obtained under the exclusion of \rcode{ComplcRF}. \rcode{BlwRF} and \rcode{MddsPLS} are again among the worst-performing methods for tebmps 2, 3, and 4. For tebmps 3 and 4 (cf.\ Supplementary Figure S25 for tebmp 4) \rcode{ImpRF} also is among the worst-performing methods, but this is seen less clearly when studying the results for each combination of trbmp and tebmp (Section \ref{sec:septrbmptebmp}) and not at all for the AUC and the accuracy (Supplementary Figures S23, S24, S26, S27, S29, and S30). An explanation for why we observed this slightly worse performance of \rcode{ImpRF} for tebmps 3 and 4 in the case of Brier could be the following: for larger numbers of blocks in the test data more blocks and thus more missing values are retained in the training data meaning that larger proportions of missing values need to be imputed.

The differences between the methods in performance were, as in the previous sections, smaller for the AUC and the accuracy (Supplementary Figures S23, S24, S26, S27, S29, and S30). Excluding \rcode{ComplcRF} and \rcode{MultisRF} (Supplementary Figures S25 to S30) did again not change the results strongly, except for tebmp 4, as discussed above. For the raw values of the performance metrics, see Supplementary Figures S31 to S39.

\subsubsection{Performance separately for each combination of trbmp and tebmp}
\label{sec:septrbmptebmp}

\begin{figure}
\centering
\includegraphics[width=0.8\textwidth]{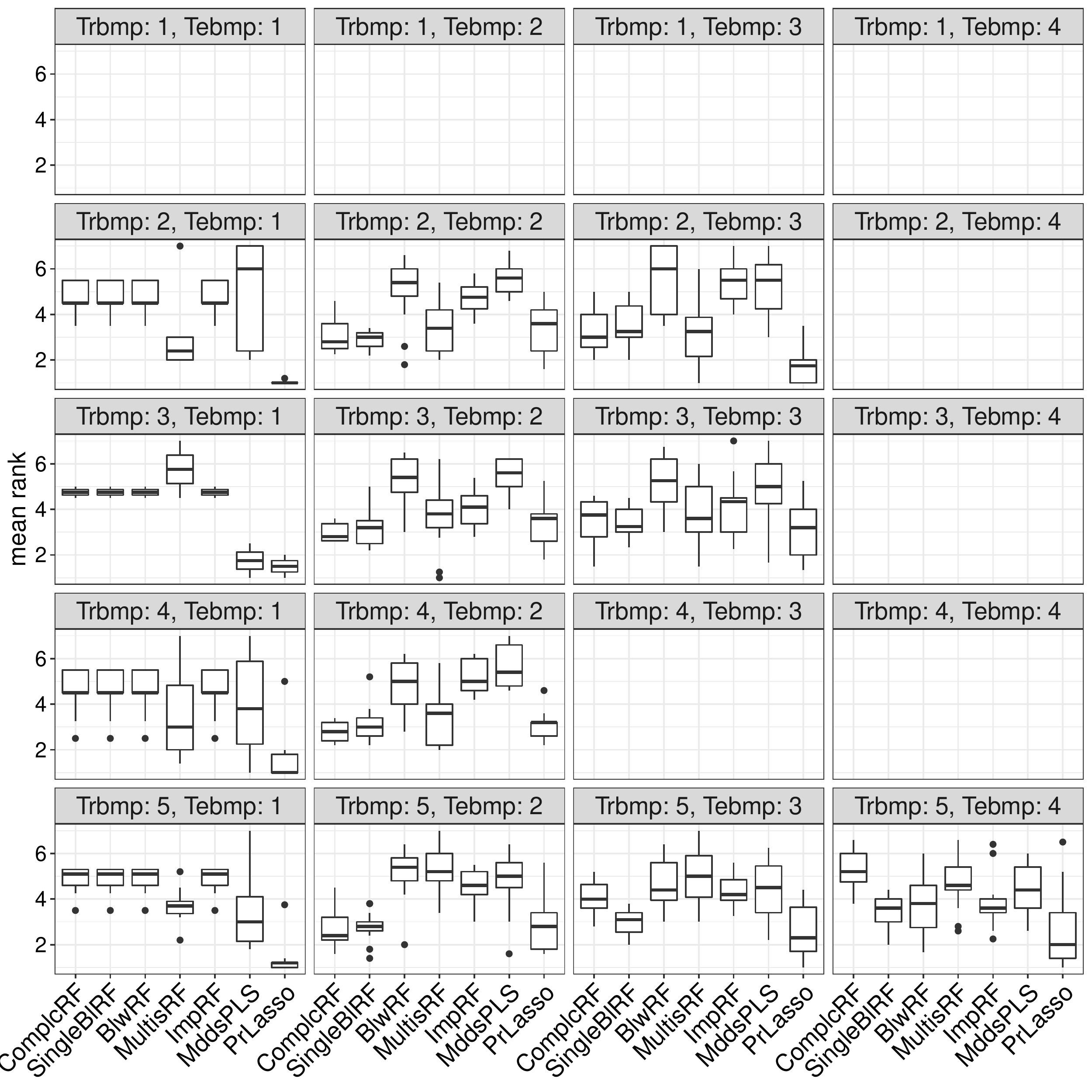}
\caption{Ranks each method achieved among the other methods in terms of the Brier score -- separately by each combination of trbmp and tebmp. The ranks were computed for each combination of tebmp, tebmp, data set, and repetition, where only those repetitions were considered for which the results were available for all seven methods. The ranks were then averaged across the repetitions. Smaller ranks indicate a better performance.}
\label{suppfig:trbmp_tebmp_Brier_ranks}
\end{figure}

In the previous two subsections we studied the performances of the methods separately by trbmp and by tebmp, but not separately by the various combinations of trbmps and tebmps. Figure \ref{suppfig:trbmp_tebmp_Brier_ranks} shows the ranks each method achieved among the other methods with respect to the Brier score separately the different combinations of trbmps and tebmps. Because some methods were not applicable for certain (combinations of) trbmps and tebmps, to interpret the results obtained for all combinations of trbmps and tebmps, we also have to consult Supplementary Figures S42 and S45, which show these results under the exclusion of \rcode{ComplcRF} and \rcode{MultisRF}, respectively.

One observation that can be made across all three performance metrics considered is that for tebmp 4, that is, the setting with no missing observations in the test data, \rcode{ComplcRF} performed much worse for trbmp 5 than for trbmp 1 (Supplementary Figures S45 to S47). This can most probably be explained by the fact that for trbmp 1 all observations are complete, whereas for trbmp 5 only the observations in the first subset are complete. This is why for trbmp 1 there are many more observations available for training in the case of \rcode{ComplcRF}. In Section \ref{sec:septrbmp}, we already made the observation that \rcode{ComplcRF} performed worse for trbmp 5, but here, when focusing on tebmp 4 this worsening is considerably stronger because if all blocks are available in test data (i.e. for tebmp 4), the complete observations are always restricted to the first subset in trbmp 5.

Another observation which can be made for the Brier score (Figure \ref{suppfig:trbmp_tebmp_Brier_ranks}) and the accuracy (Supplementary Figure S41), however, not for the AUC (Supplementary Figure S40), is that in the case of tebmp 1, \rcode{MultisRF} performed worse for trbmp 3 than for trbmps 2, 4, 5. This may be explained by the fact that, starting with the first splits, \rcode{MultisRF} prunes all trees, cutting tree branches at splits for which variables are used that do not occur in the test set. Because the clinical covariates are so few in comparison to the omics covariates it is very unlikely that the first splits use clinical covariates, which is why, if the test data only feature the clinical covariates (i.e. for tebmp 1), many trees are removed entirely from the forests, which naturally leads to a worsening of the predictive performance. Finally, the reason why \rcode{MultisRF} performed worse for trbmp 3 than for the other trbmps (2, 3, and 4) could be that for trbmp 3 more blocks are observed, which could make it particularly unlikely that the first splits use clinical covariates. 

For the combination trmbp 1 and tebmp 4 \rcode{PrLasso} did not clearly outperform \rcode{ComplcRF} (Supplementary Figures S45 to S47). This is interesting against the background that in Section \ref{sec:septebmp} it was seen that \rcode{PrLasso} outperformed \rcode{ComplcRF} (as well as \rcode{SingleBlRF}, \rcode{BlwRF}, and \rcode{ImpRF}) in the case of tebmp 1, for which only the clinical block is used as covariate data. This indicates that \rcode{PrLasso} may only outperform \rcode{ComplcRF} (that is, standard random forests) if only clinical covariate data are used, but not necessarily for multi-omics data.

The results obtained for the AUC and the accuracy were mostly very similar to those obtained for the Brier score; however, the differences between the methods' performance tended to again be smaller for the former two performance metrics (Supplementary Figures S40, S41, S43, S44, S46, and S47). With a few exceptions, excluding \rcode{ComplcRF} and \rcode{MultisRF} hardly changed the results obtained for the remaining methods (Supplementary Figures S42 to S47). The raw values of the performance metrics are shown in Supplementary Figures S48 to S56.

\section{Discussion}
\label{sec:discussion}

For the great majority of the different trbmps and tebmps we considered, \rcode{PrLasso} performed best with respect to the Brier score. In Section \ref{sec:septebmp} we obtained strong evidence that \rcode{PrLasso} better exploited the predictive information contained in the clinical covariates than the random forest-based methods. This could be an important reason for the superior performance of \rcode{PrLasso} in our comparison study, given that the clinical covariates have been found to be very important to prediction with multi-omics data \citep{Herrmann:2020}.

\rcode{BlwRF} and \rcode{MddsPLS} performed worst for most settings. These two methods have in common that they treat the blocks independently from each other, where for prediction all blocks are used. Treating the different blocks separately might not be effective because the predictive information contained in them is overlapping. Surprisingly, the naive methods were not among the worst methods in general and even among the best for some settings. In contrast, \rcode{PrLasso} demonstrated a more robust behavior in the sense that it was consistently among the best methods. As expected, the complete case approach \rcode{ComplcRF} did not perform well if there were only few complete observations in the training data.

In its current form \rcode{MultisRF} is not very well suited for the multi-omics case. An important reason for this is the pruning procedure performed prior to prediction (cf.\ Section \ref{sec:approaches}). If the test data contain blocks of small size, this procedure is expected to lead to the removal of large proportions of the trees in the random forests constructed with \rcode{MultisRF}. We saw this in our comparison study, where, as discussed in Section \ref{sec:septrbmptebmp}, the predictive performance was diminished if the test data only contained clinical covariates, but at the same time in the test data many omics blocks were available. \rcode{MultisRF} could also be better adjusted to the multi-omics case by replacing the standard random forests constructed using the different subsets by a prediction method that takes the multi-omics structure into account, for example, by the block forests method \citep{Hornung:2019}.

The results differed across the different performance metrics. For example, while \rcode{PrLasso} performed substantially better than the other methods with respect to the Brier score, there was no clear winner among the methods for the AUC. In general, great care has to be taken in the interpretation of our results. As all studies based on real data sets, ours may have yielded different results to some extent if we had considered other real data sets as a basis (other inclusion criteria, or data from other databases) and more random repetitions for each of these real data sets. Moreover, while the number of data sets is larger than in many benchmark studies that are based on few, say four or five, data sets the number of included data sets is still limited. \citet{Niessl:2022} have shown that the results of benchmark studies are in general variable and strongly affected by analytic choices even if large numbers of data sets are used. Against this background, to avoid drawing non-replicable conclusions, we took great care to offer reasonable explanations for our observations. In the cases of \rcode{ComplcRF} and \rcode{MultisRF} there were no results for certain trmpbs and tempbs. In general, systematically missing values of this kind are an issue in empirical benchmark studies, which may lead to biased results. However, by only including repetitions obtained for all seven methods in our visualisations we took care to present our results appropriately in view of this issue. Moreover, the additional figures presented in each case that excluded \rcode{ComplcRF} and \rcode{MultisRF}, respectively, suggested that our results are quite robust against biases caused by the missing values.

Our study has a number of further limitations that may be addressed in future research. For reasons of clarity we investigated only a limited number of simplified missing patterns, which may not cover all potential scenarios encountered in reality. Practical scenarios may for example feature more \lq\lq irregular patterns'' with subsets of patients of very different sizes. Moreover, there could be missing values in individual covariates, that is, missing values beyond whole types of omics data missing for subsets of patients. Importantly, in practice the data in the subsets that feature different block combinations often stem from different sources (e.g. generated at different time points or using different machines). This leads to systematic differences between the data subsets that are generally known as batch effects \citep{Li:2009}. The training data and test data subsets in our empirical study were random partitions of the same data sets and thus did not feature batch effects. In the training data such batch effects can be corrected using batch effect removal methods such as ComBat \citep{Johnson:2006}. For correcting the test data correspondingly, add-on batch effect removal can be used \citep{Hornung:2016}, where the test data is transformed to be similar to the training data in distribution. This add-on batch effect removal helps to improve the predictive performance which tends to deteriorate in the presence of batch effects \citep{hornung2017improving}.

We assumed that the missingness pattern is known in the test data before training the model. While, in principle, the model can be retrained depending on the respective test data missingness pattern before prediction, this may not always be feasible in practice. Applying models to test data with varying missingness patterns is fundamentally different from the setting considered in the present study and requires different, possibly more complex techniques.

While we claim that the choice of the TP53 mutation as a binary response variable considered as surrogate for disease outcome is acceptable from the purely methodological perspective adopted here, it may make sense to consider other response variables or to extend the study to other types of response variables such as censored survival times. However, not all methods considered in our study are capable to handle this case. 
Our study could also be extended to include further methods, in particular methods not implemented in R, which we excluded from our comparison. 

\section{Conclusion}
\label{sec:conclusions}

We first provided a state-of-the-art literature overview on prediction methods for block-wise missing multi-omics covariate data. Subsequently, we presented a large-scale benchmark comparison study of some of these methods. The results of this study may aid applied researchers confronted with block-wise missing multi-omics data to select suitable methods. Nevertheless, given the generally high variability of the findings of benchmark studies, it is important to not over-interpret details of the results of our study. In addition to applied researchers, the literature overview and the benchmark study may also aid methodological researchers in developing new, stronger methods that share the strengths of the most promising methods, while addressing their weaknesses.

\section*{Acknowledgements}
The authors thank Anna Jacob for valuable language corrections. This work was supported by the German Science Foundation [grant number HO6422/1-2 to Roman Hornung; grant numbers BO3139/6-2 and BO3139/4-3 to Anne-Laure Boulesteix].

\section*{Supplementary Materials and R code}
\begin{sloppypar}
The Supplementary Materials can be found at the following link: \url{http://www.ibe.med.uni-muenchen.de/organisation/mitarbeiter/070_drittmittel/hornung/missmultiomics_suppfiles/suppmat_hornungetal2023.pdf}. The data that support the findings of this study are openly available in figshare at \url{http://doi.org/10.6084/m9.figshare.20060201.v1}. All R code written to perform and evaluate the analyses, including data pre-processing, is available on GitHub (\url{https://github.com/RomanHornung/bwm_article}).
\end{sloppypar}

\nocite{*}

\end{document}